\documentclass[%
reprint,
superscriptaddress,
amsmath,amssymb,
aps,
prl,
]{revtex4-1}
\usepackage{soul} 
\usepackage{xcolor} 
\usepackage{graphicx}
\usepackage{dcolumn}
\usepackage{bm}
\usepackage{amsmath}
\usepackage{hyperref}
\hypersetup{colorlinks=true,linkcolor=blue,anchorcolor=blue,citecolor=blue,urlcolor=blue}
\usepackage{lineno}
\usepackage{enumerate}
\usepackage{color}
\usepackage{microtype}

\begin{document}
\preprint{APS/123-QED}

\title{Bridging the Kinetic-Fluid Gap: Ion-Driven Magnetogenesis to Prime Cosmic Dynamos}

\author{X. Liu}
\affiliation{ 
State Key Laboratory of Dark Matter Physics, Key Laboratory for Laser Plasmas, School of Physics and Astronomy, Shanghai Jiao Tong University, Shanghai 200240, China
}%
\affiliation{
Collaborative Innovation Center of IFSA (CICIFSA), Shanghai Jiao Tong University, Shanghai 200240, China
}

\author{D. Wu}
\email{dwu.phys@sjtu.edu.cn}
\affiliation{ 
State Key Laboratory of Dark Matter Physics, Key Laboratory for Laser Plasmas, School of Physics and Astronomy, Shanghai Jiao Tong University, Shanghai 200240, China
}%
\affiliation{
Collaborative Innovation Center of IFSA (CICIFSA), Shanghai Jiao Tong University, Shanghai 200240, China
}

\author{J. Zhang}
\email{jzhang@iphy.ac.cn}
\affiliation{ 
State Key Laboratory of Dark Matter Physics, Key Laboratory for Laser Plasmas, School of Physics and Astronomy, Shanghai Jiao Tong University, Shanghai 200240, China
}%
\affiliation{
Collaborative Innovation Center of IFSA (CICIFSA), Shanghai Jiao Tong University, Shanghai 200240, China
}
\affiliation{
\mbox{Institute of Physics, Chinese Academy of Sciences, Beijing 100190, China}}

\date{\today}

\begin{abstract}
The origin of cosmic magnetic fields is widely attributed to the amplification of weak seed fields by turbulent dynamos. However, a critical understanding gap remains between the microscopic generation of these seeds and the macroscopic onset of the dynamo. Current kinetic models, often constrained to electron scales, predict premature saturation via magnetic trapping, leaving the generated fields potentially too weak and small-scale to effectively prime magnetohydrodynamic (MHD) processes. Here, using high-resolution kinetic simulations with a realistic mass ratio, we reveal the physics of this unexplored ion-kinetic regime. Under generalized continuous shear driving, used to simulate ubiquitous macroscopic flows, we demonstrate that the saturation of electron instabilities is not the endpoint but a precursor to a distinct, ion-dominated evolution. Massive ions, sustaining the velocity shear, trigger a subsequent filamentation instability that accesses the vast ion kinetic energy reservoir. This mechanism amplifies the magnetic energy by orders of magnitude beyond the electron-saturation limit, expanding the field coherence to ion scales. Our results establish ion kinetics as the essential ``missing link'' that bridges the divide between microscopic plasma instabilities and macroscopic cosmic dynamos.
\end{abstract}

\maketitle
$ Introduction $.---
As the long-range interaction capable of shaping large-scale structures, magnetic fields play a pivotal role in critical astrophysical processes, including accretion disk dynamics \cite{Balbus1991, Balbus1998, Hoshino2015}, star formation \cite{McKee2007, Shu1987, Crutcher2012}, supernovae \cite{LeBlanc1970, Mosta2015, Janka2012}, and cosmic ray acceleration \cite{Bell1978, Blandford1987}. Observations indicate that magnetic fields with strengths on the order of $\mu$G are ubiquitous in the intracluster medium (ICM) \cite{Carilli2002, Beck2016, vanWeeren2019}. The prevailing scenario attributes the origin of these macroscopic fields to weak ``seed'' fields: once generated, these seeds undergo an inverse cascade process---characterized by the decay of magnetic energy accompanied by the growth of the coherence scale---and are subsequently amplified exponentially by turbulent dynamos until reaching saturation \cite{Banerjee2004, Schekochihin2004, Brandenburg2005, Zhou2023}. Consequently, identifying a mechanism capable of generating seed fields with sufficient intensity and the potential for large-scale coherence is a prerequisite for resolving the problem of cosmic magnetogenesis \cite{kulsrud2008, widrow2012, subramanian2016, pudritz1989, subramanian1994, gnedin2000}.

Kinetic instabilities in collisionless plasmas, notably the Weibel or filamentation instability, are considered a potent mechanism for the generation of seed magnetic fields \cite{weibel1959, Fried1959, Davidson1972, medvedev1999, silva2003}. By converting particle free energy into magnetic energy, these instabilities rapidly establish small-scale magnetic fields. The growth of magnetic fields saturates through electron trapping once the electron Larmor radius becomes comparable to or smaller than the characteristic length scale of the field \cite{Davidson1972, medvedev1999, Kato2005}, eventually resulting in field decay.
This electron-trapping paradigm has long dominated the understanding of microscopic magnetogenesis, implying that the intensity of seed magnetic fields is strictly constrained by electron kinetics \cite{zhou2022, Pucci2021}.

However, this classical picture overlooks the crucial impact of the realistic ion-electron mass ratio in the cosmic environment. In electron-positron (pair) plasmas, the symmetric kinetic response results in a lack of potent driving mechanisms after saturation at electron scales, hindering further amplification toward the turbulent dynamo regime. In contrast, realistic ion-electron plasmas introduce significant spatiotemporal scale separation due to the vast mass disparity. 
Such conditions are particularly prevalent in supersonic shear flows associated with cold-stream accretion onto galaxies \cite{dekel2009}, the collapse of large-scale structures \cite{ryu2008}, and the outskirts of accretion disks \cite{sharma2006}. These systems often possess extremely high Mach numbers, allowing kinetic instabilities to tap into the vast reservoir of directed bulk kinetic energy before the plasma is thermalized by shocks to high temperatures ($\sim$keV).
This separation of scales raises a critical yet largely unexplored question: whether and how the ions can govern the post-saturation dynamics to sustain magnetic amplification, bridging the gap between weak electron-scale fields and the field strengths required by turbulent dynamos.

In this Letter, we address this problem by conducting high-resolution kinetic simulations with a realistic proton-to-electron mass ratio, under a generalized continuous shear driving framework. We reveal that the realistic mass disparity breaks the standard electron-saturation barrier. Specifically, the large ion inertia delays thermal free-streaming and decouples ion dynamics from electron-scale magnetic trapping, allowing the velocity shear to persist. Consequently, a distinct ion-driven filamentation instability is excited, accessing a free-energy reservoir inaccessible to pair plasmas. This ion-driven mechanism supersedes the saturated electron dynamics, driving magnetic amplification orders of magnitude beyond the pair-plasma limit and expanding the coherence length to ion scales. Our results demonstrate that this ion-mediated phase is the essential link priming the cosmic dynamo.

$Methods$.---In astrophysical systems, the macroscopic background flows are ubiquitous and usually immense. Therefore, we introduce an external forcing to mimic the continuous shear driving \cite{zhou2022}. To investigate the impact of ion kinetics on seed magnetic field generation, we performed fully kinetic simulations with the high-order implicit particle-in-cell (PIC) code LAPINS \cite{WuPoP2014,WuPoP2015,WuPRE2017stopping,WuPRE2017ionization,WuHPLSE2018,WuPRE2019,WuAIPAdv2021}. A proton-electron system with a realistic mass ratio $m_{\rm{i}}/m_{\rm{e}} = 1836$ is initialized uniformly in the $z$-$y$ plane. The simulation domain size is set to $L=1600\Delta z=20d_{\rm{i}} \approx 858d_{\rm{e}}$. This yields a grid resolution $\Delta z \approx 0.536d_{\rm{e}}$, ensuring that both electron and ion kinetic scales are well resolved, where the skin depth for species $\rm{s}$ is defined as $d_{\rm{s}} \equiv c/\omega_{\rm{p,s}}$. Periodic boundary conditions are applied in all directions, and the number of macro-particles per cell (ppc) is set to 100. Both species are initialized with Maxwellian velocity distributions. The initial electron temperature is $\theta_{\rm{e}} \equiv T_{\rm{e}}/m_{\rm{e}} c^2=1/16$, corresponding to a sub-relativistic thermal velocity $v_{\rm{th,e}} \equiv \sqrt{T_{\rm{e}}/m_{\rm{e}} } = 0.25 c$. To reduce the computational cost, the initial ion temperature is boosted to $\theta_{\rm{i}}=25\theta_{\rm{e}}$, resulting in an ion thermal velocity $v_{\rm{th,i}} \equiv \sqrt{T_{\rm{i}}/m_{\rm{i}} } \approx 0.03 c$. We drive the shear flow by applying an external sinusoidal force field $\mathbf{F}_{\rm{ext,s}}=m_{\rm{s}} a_0 \sin(2\pi z/L)\hat{\mathbf{e}}_{y}$. The driving amplitude is defined as $a_0=S_0 v_{\rm{th,i}} v_{\rm{th,e}}/L$ to ensure sufficient kinetic excitation for both species, with the strength parameter $S_0$ set to $0.5\pi^2$. Driven by the applied force, the system exhibits distinct temporal scale separation in its kinetic evolution, stemming from the large mass disparity between electrons and ions. To elucidate this cross-scale physical process, we discuss the simulation results in three consecutive phases: the initial unmagnetized stage, the electron-dominated phase, and the ion-dominated phase.

\begin{figure}[h]
  \includegraphics[scale=0.55]{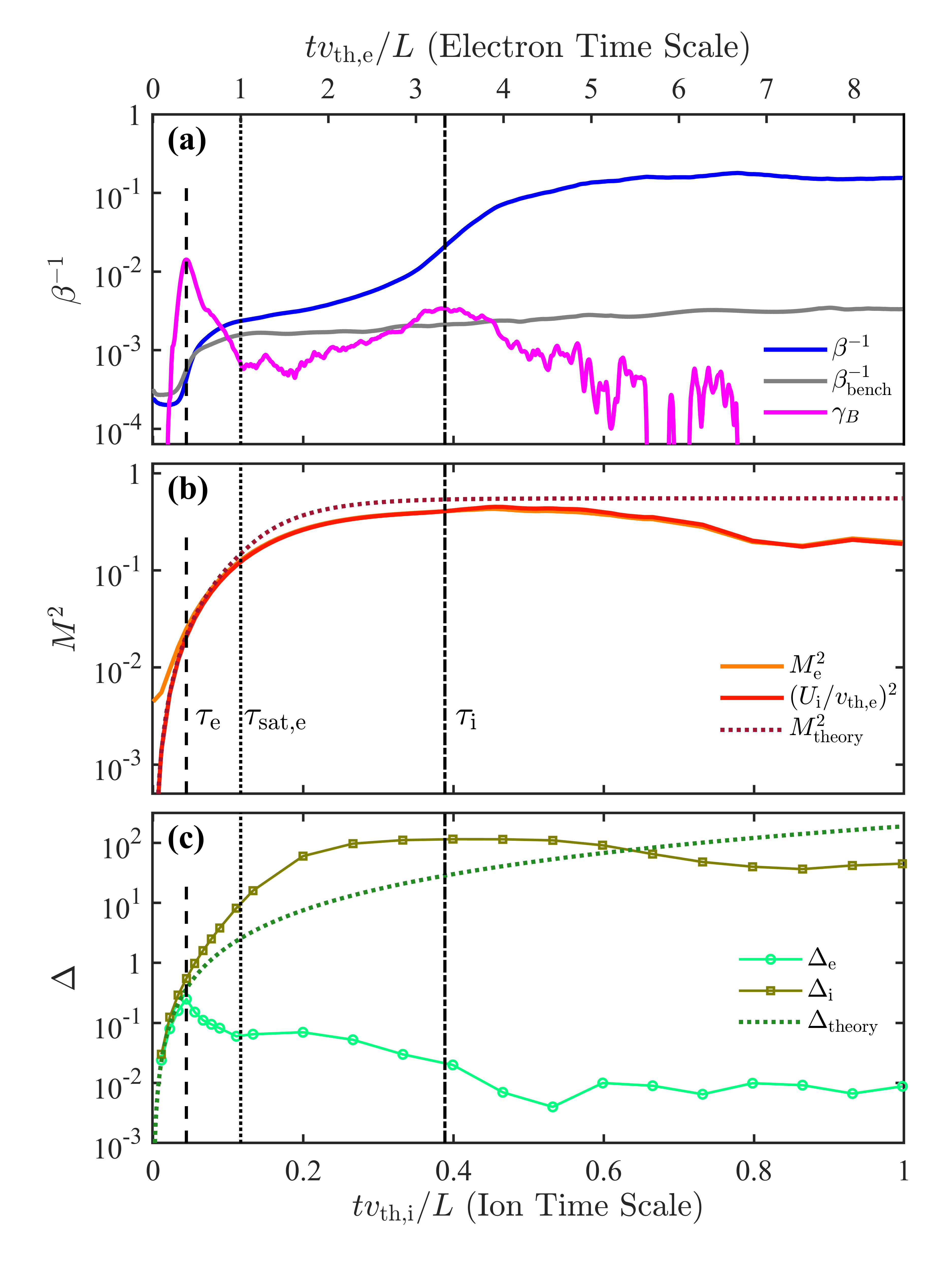}
    \caption{Temporal evolution of various parameters. Colored dashed lines represent theoretical values for the unmagnetized stage; gray lines represent the magnetic field evolution for the electron-positron pair case. Black vertical dashed lines indicate three typical moments. }
    \label{fig:1}
\end{figure}

$ Unmagnetized\ Stage $.---In the initial unmagnetized stage, electromagnetic fields are negligible, and the system's evolution can be described by the Mach number and the anisotropy parameter. The macroscopic motion is characterized by the Mach number, $M_{\rm{s}} \equiv \sqrt{\langle U_{\rm{s}}(t, \mathbf{x})^2 \rangle} / v_{\rm{th,s}} = \sqrt{\langle (\int \mathrm{d}^3\mathbf{v}\, \mathbf{v} f_{\rm{s}})^2 \rangle} / v_{\rm{th,s}}$, where $f_{\rm{s}}(t, \mathbf{x})$ is the distribution function for species $\rm{s}$, and $\langle \cdot \rangle$ denotes spatial averaging. The thermal properties are given by the thermal pressure tensor $\mathbf{P}_{\rm{s}}(t, \mathbf{x}) \equiv \int \mathrm{d}^3\mathbf{v}\, m_{\rm{s}} (\mathbf{v}-\mathbf{U}_{\rm{s}})(\mathbf{v}-\mathbf{U}_{\rm{s}}) f_{\rm{s}}$. The degree of anisotropy is quantified by $\Delta_{\rm{s}}(t) \equiv \sqrt{\langle (P_{\rm{max,s}}/P_{\perp,\rm{s}})^2 \rangle} - 1$, where $P_{\rm{max,s}}$ is the maximum eigenvalue of the local pressure tensor $\mathbf{P}_{\rm{s}}$, and $P_{\perp,\rm{s}}$ is the average of the other two eigenvalues perpendicular to the principal direction. In the limit of $\epsilon_{\rm s}=t v_{\rm{th,s}}/L \ll 1$, the Mach number and anisotropy parameter can be derived theoretically. With the normalized acceleration $\hat{a}_{0,s} = a_0 L / v_{\rm{th,s}}^2$, we have \cite{SupplementalMaterial}:
\begin{eqnarray}
    M_{\rm{s}}(t) &=& \frac{1}{\sqrt{2}} \hat{a}_{0,\rm{s}} \frac{t v_{\rm{th,s}}}{L} + \mathcal{O}(\epsilon_{\rm s}^3), \label{eq:Ms_growth}\\
    \Delta_{\rm{s}}(t) &=& \frac{3\pi}{2\sqrt{2}} \hat{a}_{0,\rm{s}} \left(\frac{t v_{\rm{th,s}}}{L}\right)^2 + \mathcal{O}(\epsilon_{\rm s}^3). \label{eq:Delta_growth}
\end{eqnarray}

On macroscopic fluid scales, electrostatic fields effectively couple electron and proton dynamics, forcing both species to maintain synchronized mean flows and identical anisotropy evolution.
As reflected in Eqs. (\ref{eq:Ms_growth}) and (\ref{eq:Delta_growth}), the identical evolution of mean fluid velocities ($U_{\rm s}=M_{\rm s}v_{\rm th,s}$) and anisotropy arises because the lower thermal velocities of heavier species are compensated by slower mixing processes.
In this unmagnetized phase, particles gradually form shear flows under the external forcing. Due to free-streaming in the $z$-direction, particles develop anisotropy in phase space. This velocity anisotropy constitutes the free energy source driving the microscopic instability, which subsequently converts this free energy into magnetic energy, realizing spontaneous magnetic field generation and amplification.

\begin{figure}[t]
  \includegraphics[scale=0.16]{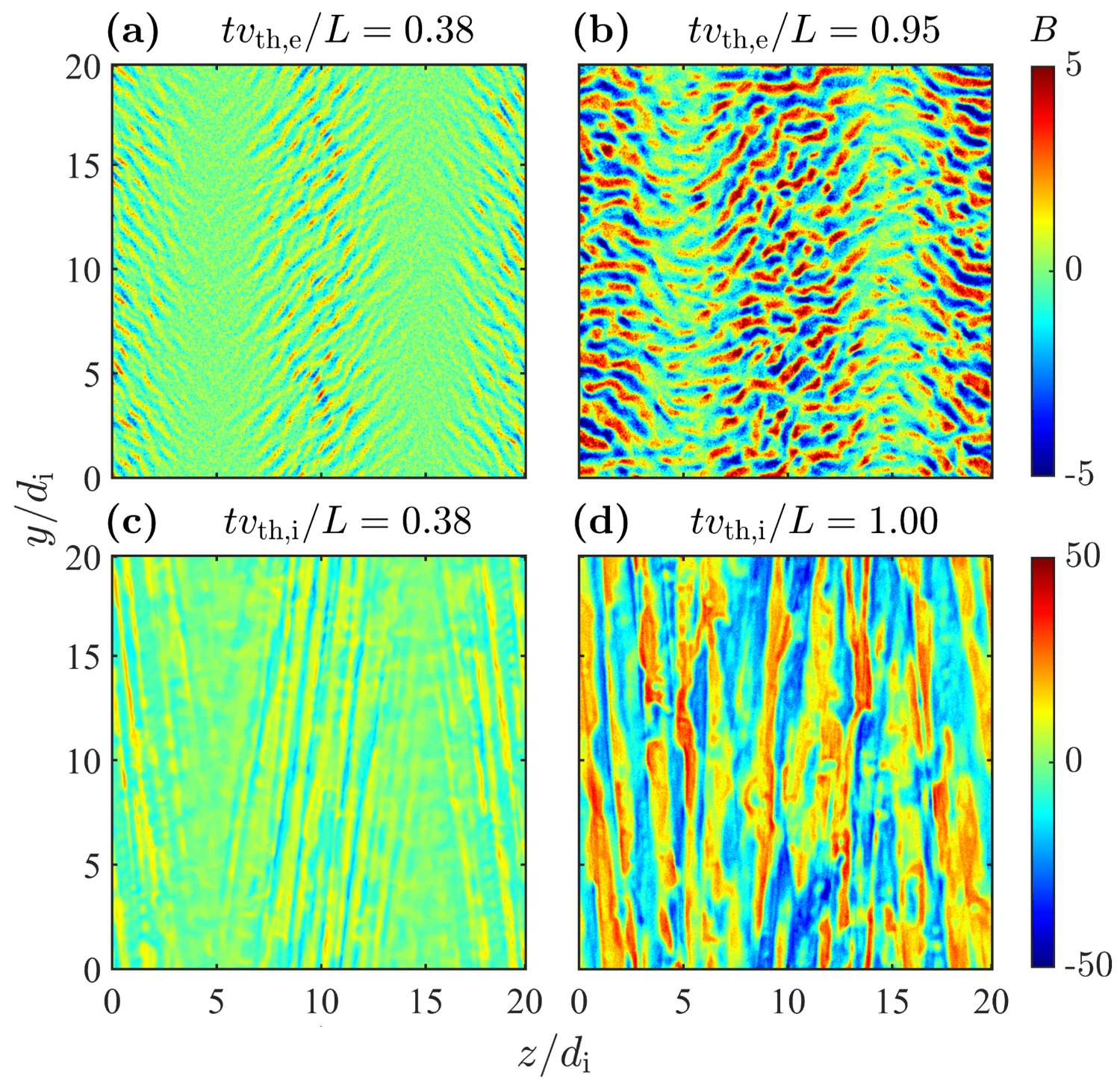}
    \caption{Magnetic field distributions at $\tau_{\rm{e}}$, $\tau_{\rm{sat,e}}$, $\tau_{\rm{i}}$, and the final simulation moment. }
    \label{fig:2}
\end{figure}

The spatial free-streaming of particles in the $z$-direction eventually leads to the saturation of the macroscopic fluid velocity growth. When the free-streaming distance in $z$ becomes sufficiently large, particles experience different phases of the shear field. The acceleration and deceleration effects applied by the external field cancel each other out, halting the growth of the macroscopic flow velocity and causing the Mach number to saturate. The time scale required is approximately $L/(2\pi v_{\rm{th,s}})$, yielding a saturation value for the particle Mach number of:
\begin{equation}
    M_{\rm{s}}^{\rm{sat}} \approx \frac{a_0 L}{2\pi v_{\rm{th,s}}^2}.
    \label{eq:Ms_sat}
\end{equation}

Microscopic charge separation fields couple the electron dynamics to the massive protons, allowing electrons to transcend their individual kinetic limits and reach much higher saturation Mach numbers. While Eq.~(\ref{eq:Ms_sat}) predicts a significantly lower saturation threshold for electrons ($M_{\rm e}^2 \sim 0.01$), Fig.~\ref{fig:1} demonstrates that they track the proton motion until the proton-dominated saturation threshold (horizontal dashed line) is reached. To highlight this synchronization, the proton curve in Fig.~\ref{fig:1} is normalized by the electron thermal velocity. Consequently, the system’s macroscopic flow is no longer constrained by electron kinematics but is governed by the larger proton inertia, which is the first fundamental distinction from symmetric electron-positron pair systems.

\begin{figure}[t]
  \includegraphics[scale=0.16]{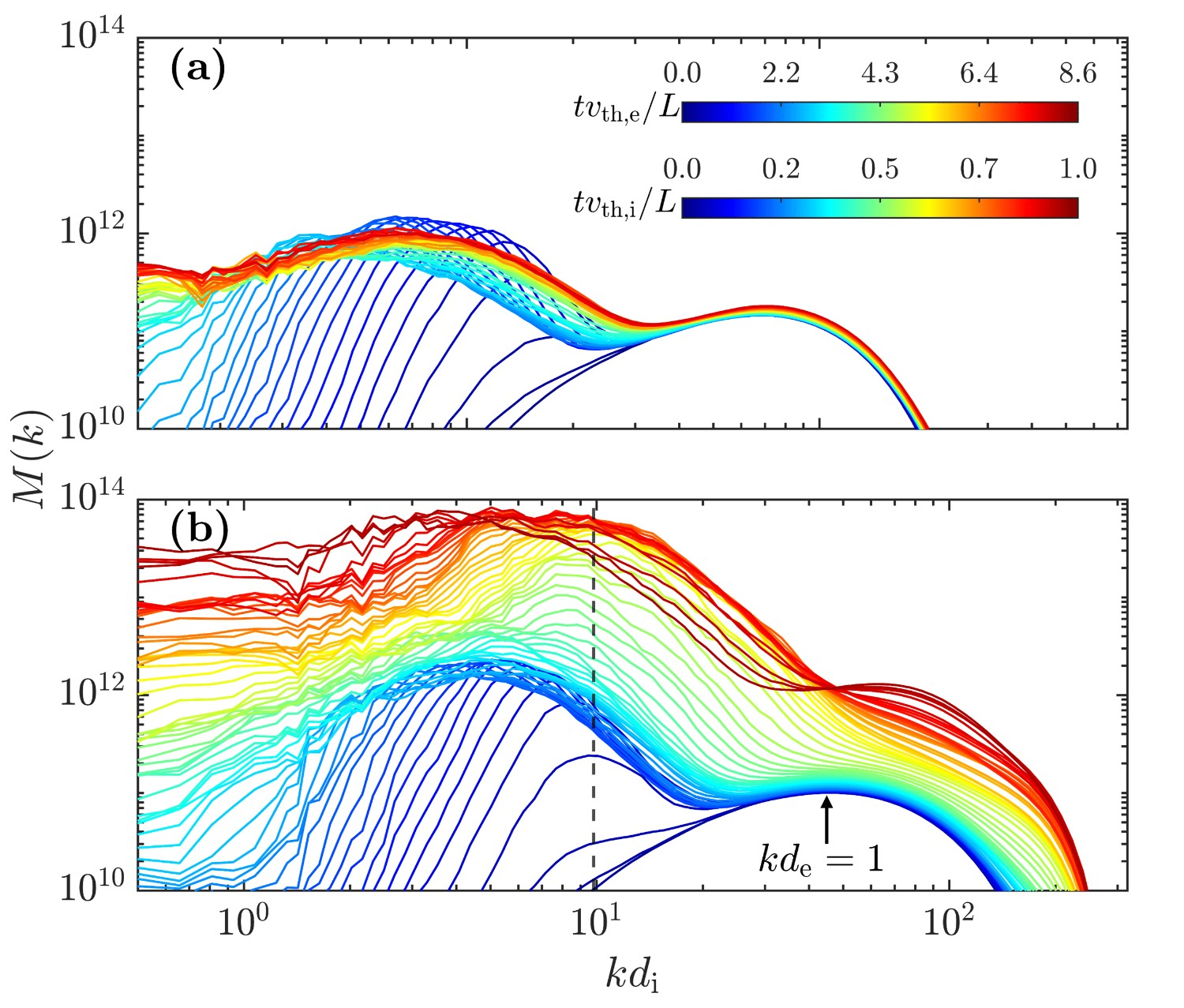}
    \caption{Evolution of magnetic energy spectra for (a) the electron-positron pair system and (b) the proton-electron system. The vertical dashed line in (b) indicates the theoretical wave number of maximum growth rate. }
    \label{fig:3}
\end{figure}

$ Electron-Dominated\ Phase $.---
The electron Weibel instability constitutes the initial growth phase of the magnetic field. As shown in Fig.~\ref{fig:1}, at $\tau_{\rm{e}} \sim 0.4 L/v_{\rm{th,e}}$, both the growth rate of the magnetic energy $\beta^{-1} \equiv \langle B^2/8\pi \rangle / \langle m_{\rm{e}} n_{\rm{e}} v_{\rm{th,e}}^2 \rangle$ and the electron anisotropy parameter $\Delta_{\rm{e}}$ reach their maximums. Subsequently, the electron anisotropy decreases, indicating the conversion of electron free energy into magnetic energy, while proton anisotropy continues to rise. For comparison, the gray line shows the magnetic energy evolution for an electron-positron pair system under identical conditions (except for the mass ratio). The onset time of the electron Weibel instability in the pair system is consistent with that in the proton-electron system. This confirms that the initial magnetic energy growth in the proton-electron system is dominated by electrons.

Electron Weibel fields first appear at the locations of maximum shear ($z=0, L/2$ and $L$), and subsequently evolve toward larger spatial scales through merging, as shown in Fig.~\ref{fig:2}(a) and (b). A more in-depth analysis of the magnetic energy spectrum is conducted in Fig.~\ref{fig:3}. Despite the initial spectral peaks at $k \sim 1/d_{\rm e}$ arising from thermal noise, a new and higher peak subsequently emerges in the proton-electron system at $k \sim 0.23/d_{\rm e} \sim 10/d_{\rm i}$, aligning with the theoretical wavenumber of maximum growth rate (black vertical line). A similar peak also appears at a comparable location in the pair system. Thereafter, in both systems, this peak continuously shifts to the left, consistent with the merging of magnetic fields toward larger scales seen in Fig.~\ref{fig:2}. Finally, the magnetic energy spectrum of the pair system stabilizes in a ``double-hump'' shape composed of the initial fluctuation peak and the electron Weibel magnetic field peak, which is identical to the evolution of the proton-electron system on electron time scales ($t v_{\rm{th,e}}/L \le 2$).

\begin{figure}[t]
  \includegraphics[scale=0.53]{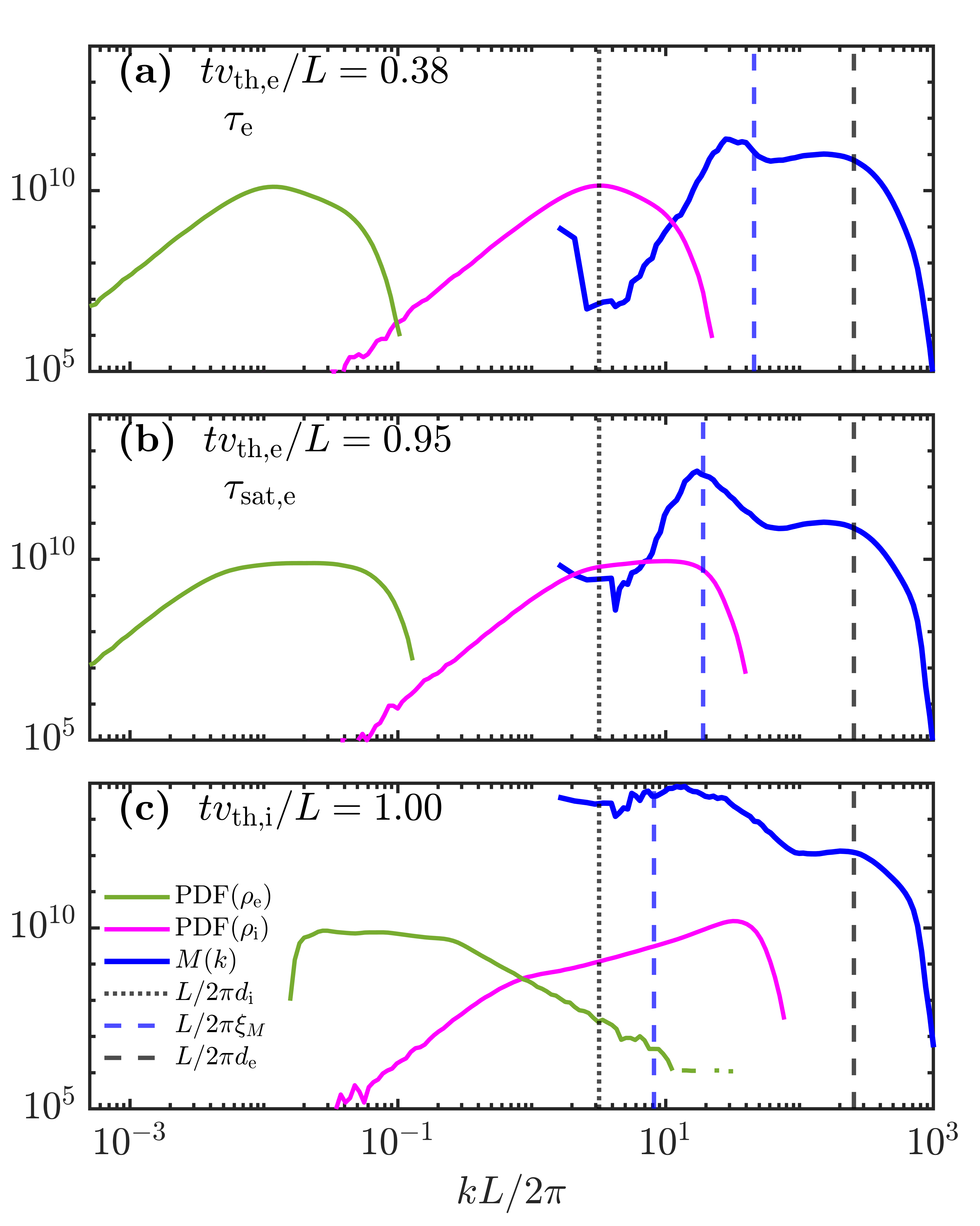}
    \caption{Magnetic energy spectra (blue solid line) and Larmor radius distributions for electrons (pink) and ions (green) at (a) $\tau_{\rm{e}}$, (b) $\tau_{\rm{sat,e}}$ and (c) the final simulation moment. Black dashed lines indicate the characteristic scales of electrons and ions, respectively. The blue vertical dashed line represents the mean magnetic field scale $\xi_M$.}
    \label{fig:4}
\end{figure}

At $\tau_{\rm{sat,e}} \sim L/v_{\rm{th,e}}$, the magnetic energy growth rate in the proton-electron system undergoes a significant decline, which represents a stage of saturation. In the pair system, magnetic energy also nearly ceases to increase after $\tau_{\mathrm{sat,e}}$. The saturation value of the electron Weibel field is given by the electron trapping condition, i.e., the product of the characteristic magnetic wavenumber $k_B$ and the electron Larmor radius $\rho_{\rm{e}}$ reaches $k_B \rho_{\rm{e}} \sim 1$. Fig.~\ref{fig:4} presents the magnetic energy spectra and the Larmor radius distributions for electrons and ions at $\tau_{\rm{e}}$, $\tau_{\rm{sat,e}}$, and the final simulation moment. The blue dashed line indicates the mean scale of the magnetic field, $\xi_M = \int \mathrm{d}k\, k^{-1} M(k) / \int \mathrm{d}k\, M(k)$. As shown in Fig.~\ref{fig:4}(b), at $\tau_{\rm{sat,e}}$, the Larmor radii of a fraction of the electrons are already smaller than the magnetic field scale, indicating electron trapping. However, due to the much larger mass of protons, the proton Larmor radius remains larger than the magnetic field scale until the end of the simulation. This explains why magnetic energy saturates in pair plasmas yet exhibits sustained growth in ion-electron plasmas, marking the second key distinction between the two systems.

$Ion-Dominated\ Phase$.---
As the system evolves into the proton time scale, protons exhibit kinetic behavior distinct from electrons. Crucially, the saturated electron Weibel fields are too weak and small-scale to deflect massive ions. Consequently, protons evolve largely unaffected by the electron-stage saturation \cite{SupplementalMaterial}. Driven by the shear force, the counter-propagating proton streams originating from maximum acceleration regions ($z=L/4$ and $3L/4$) penetrate the central region at $t \sim L/4v_{\rm{th,i}}$ via thermal free-streaming. This spatial mixing naturally leads to a double-peak distribution in momentum space as shown in Fig.~\ref{fig:5}. As detailed in the Supplemental Material, this non-thermal feature is a robust kinetic outcome of phase mixing rather than an artifact of external forcing, effectively priming the system for the ion filamentation instability, which constitutes the third key feature distinguishing the proton-electron plasma from the pair system.

The evolution of the magnetic spectrum indicates that the ion instability does not merely continue the inverse cascade of the electron magnetic field but effectively reignites the growth at ion scales. As shown in Fig.~\ref{fig:3}(b), unlike the stagnation observed in the pair system, the magnetic energy spectrum of the proton-electron system exhibits a significant overall elevation as the ion filamentation instability takes over. According to linear theory, based on the velocity and density ratios of the proton streams at different locations, the instability wavenumber is around $k d_{\rm i} \lesssim 10$. Notably, this new spectral peak grows independently at ion scales rather than shifting smoothly from the electron spectral peak. This discontinuity reveals that, despite synchronized macroscopic flows, the saturated electrons and energetic ions have kinetically decoupled. As evident in Fig.~\ref{fig:2}(b), the onset of ion instability marks a fundamental regime shift: by tapping into the immense free energy of proton anisotropy, the system effectively resets the inverse cascade, driving magnetic field evolution with renewed power at ion scales.

The spatial expansion of the proton filamentation instability drives an exponential growth in magnetic energy, eventually reaching a sub-equipartition saturation state. As the proton mixing region enlarges, new regions continuously undergo filamentation instability, causing the magnetic energy to grow exponentially. After $t \sim L/2v_{\rm{th,i}}$, the proton mixing region expands to cover the entire space, indicating that the filamentation instability has fully developed. Under the effect of mixing, the proton momentum distribution in the $y$-direction rapidly concentrates from a double-peak distribution toward $p_y=0$. With the consumption of proton free energy, the proton anisotropy parameter and Mach number decrease, and the magnetic energy gradually tends toward saturation. Our simulation results show that the magnetic energy at saturation is approximately two orders of magnitude higher than the electron saturation value, reaching a sub-equipartition level, $\beta^{-1} \sim \eta m_{\rm{i}} n_{\rm{i}} v_{\rm{th,i}}^{\rm{sat}} / m_{\rm{e}} n_{\rm{e}} v_{\rm{th,e}}^2 \sim \eta m_{\rm{i}}/m_{\rm{e}}$, where $\eta \sim 2 \times 10^{-4}$. Crucially, this substantial pre-magnetization level implies that the generated fields are strong enough to decouple from electron-scale damping, providing a robust, ion-scale seed essential for initiating the macroscopic turbulent dynamo.

$Conclusion\ and\ Discussion $.---
In this Letter, using fully kinetic simulations, we have revealed the complete physical picture of seed magnetic field generation in collisionless plasmas with a realistic proton-electron mass ratio under a generalized continuous shear driving framework. By directly comparing with electron-positron pair case, we identified how mass asymmetry breaks the classical electron saturation limit through three progressive kinetic mechanisms. First, the lower thermal free-streaming speed of protons allows them to maintain a saturation flow velocity far higher than that of electrons in the shear field, thereby accumulating immense ion free energy. Second, the mass difference leads to a decoupling of saturation mechanisms. The rapid magnetic trapping of electrons cannot constrain the motion of massive ions, allowing the proton-electron system to surpass the saturation stagnation point of the pair system. Finally, the mixing of ion streams excites the filamentation instability. This potent ion mechanism successfully takes over the weak evolution channel originally dominated by electrons, transferring magnetic field generation to ion kinetic scales and driving a significant secondary growth.

For computational feasibility, this study employed a relatively high proton thermal velocity. In realistic cosmic environments ($T_{\rm{i}} \sim T_{\rm{e}}$), the suppressed proton thermal free-streaming would likely sustain sharper velocity shears, potentially exciting stronger instabilities. Furthermore, the extent to which saturated electron-scale fields influence ion dynamics remains an open question.
While our simulations suggest kinetic decoupling, long-term feedback from electron-scale fields on ion isotropization in different regimes remains an avenue for future study.
Finally, regarding dimensionality, while 2D simulations may overestimate saturation levels by suppressing the kink mode, the core physical mechanism—ion inertia breaking through electron trapping—is robust and independent of geometric constraints.

\begin{figure}[t]
  \includegraphics[scale=0.97]{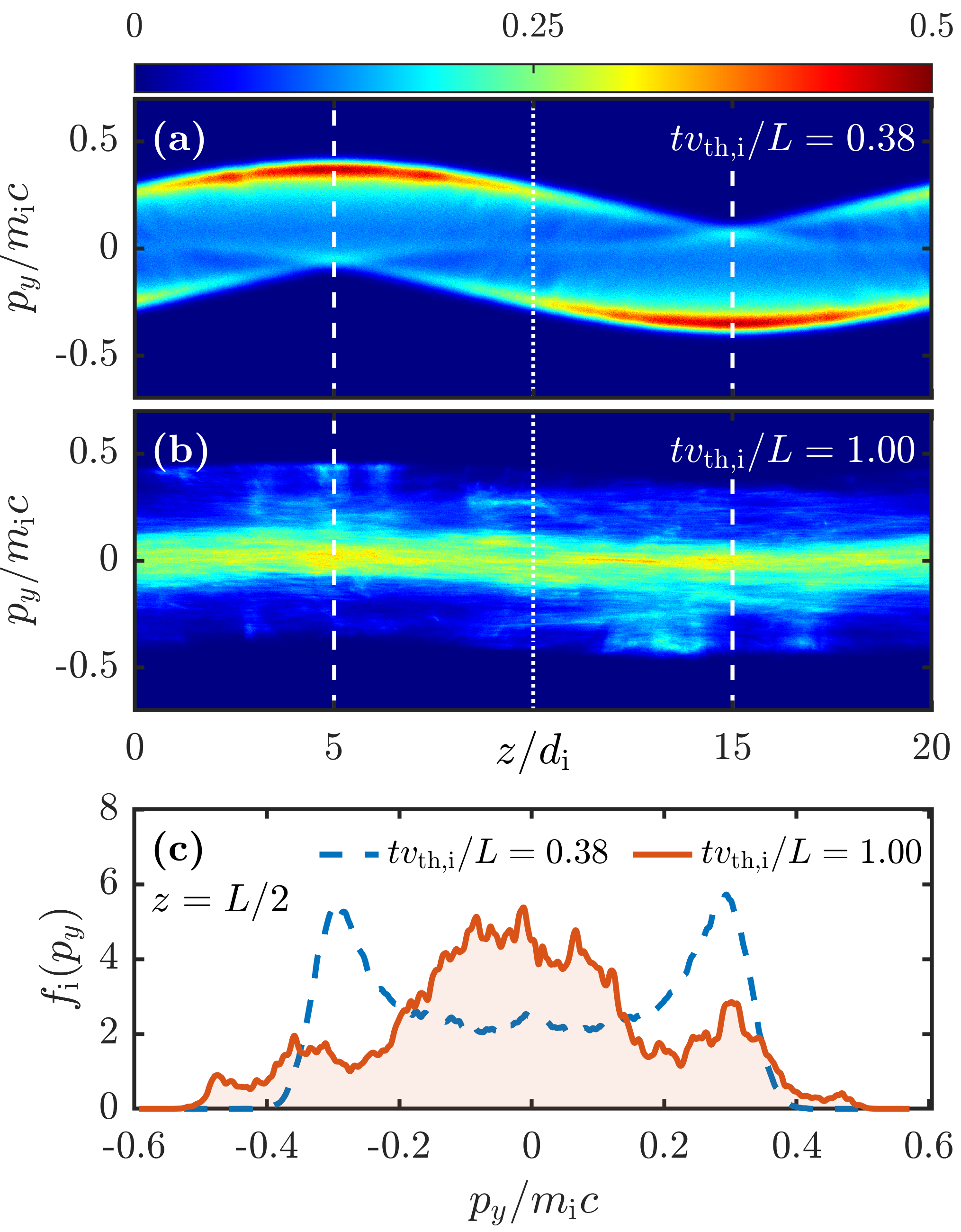}
    \caption{(a)(b) Proton phase space distributions at different moments; white dashed lines indicate the locations of the maximum acceleration region and the intermediate region. (c) Proton momentum distribution at $z=L/2$.}
    \label{fig:5}
\end{figure}

By establishing ion kinetics as the essential ``missing link,'' our results bridge the critical gap between microscopic seed generation and the onset of macroscopic turbulent dynamos. This ion-driven mechanism ensures that the ``pre-magnetized'' medium possesses sufficient strength and ion-scale coherence to effectively prime MHD processes. Consequently, this study not only revises the initial conditions for dynamo theories but also provides a robust explanation for the rapid emergence of large-scale magnetic fields in the cosmos.

\nocite{*}
\bibliography{Reference}

\end{document}